\documentclass[conference,a4paper]{IEEEtran}
\usepackage{xcolor}
\usepackage{balance}

\newsavebox{\figbox}

\ifCLASSINFOpdf
  \usepackage[pdftex]{graphicx}
\else
  \usepackage[dvips]{graphicx}
\fi
\ifCLASSOPTIONcompsoc
 \usepackage[caption=false,font=normalsize,labelfont=sf,textfont=sf]{subfig}
\else
 \usepackage[caption=false,font=footnotesize]{subfig}
\fi
\usepackage[T1]{fontenc}
\usepackage[utf8]{inputenc}
\usepackage{graphicx}
\usepackage{amsmath,amssymb,bm}
\usepackage{booktabs}
\usepackage{array}
\usepackage{cite}
\usepackage{url}
\begin{document}
%
\title{Sensitivity and Identifiability of Tumor-Related Changes in Longitudinal Microwave Breast Imaging}

\author{\IEEEauthorblockN{
Somayeh B. Shafiei,   
Ian Jeffrey,
Vahab Khoshdel,   
Joe LoVetri,    
}                                     
\IEEEauthorblockA{
Department of Electrical and Computer Engineering, University of Manitoba, Winnipeg, Canada}
 \IEEEauthorblockA{ \emph{Joe.LoVetri@UManitoba.Ca} }
}
    \vspace{-8pt}

\maketitle

\begin{abstract}

Microwave imaging (MWI) is a candidate for repeated, non-ionizing monitoring of breast tumor response to neoadjuvant chemotherapy, but between-visit changes in breast positioning and normal-tissue properties may confound tumor-related changes in the measured signals. We present a patient-specific computational framework to assess whether tumor dielectric changes are distinguishable from these competing effects. Breast, fibroglandular tissue (FGT), and tumor were segmented from longitudinal I-SPY 1 MRI of two patients at four treatment time points and converted to dielectric models. For each of the eight models, 24-port S-parameters were simulated from 1.0 to 3.0~GHz in a quasi-resonant imaging chamber, and finite-difference Jacobians were computed for six parameters: tumor permittivity and loss ($\pm$10\%), FGT permittivity ($\pm$10\%), breast rotation ($\pm$2$^\circ$), and lateral and superior--inferior translation ($\pm$1~mm). Sensitivity, distinguishability, and identifiability were quantified using baseline-normalized S-parameter changes, cosine similarity between tumor permittivity and other Jacobian columns, and singular-value analysis of the multi-frequency Jacobian. Positioning perturbations produced larger S-parameter changes than the tumor dielectric perturbations, particularly above 2~GHz. Nevertheless, tumor permittivity sensitivity patterns were weakly aligned with positioning responses, and the frequency-stacked Jacobians remained well-conditioned (condition number $\leq 2.5$) at all time points. Tumor--FGT similarity was patient-dependent, reaching 0.82 near 2~GHz in the larger-tumor case. These results indicate that tumor-related responses can remain distinguishable from positioning effects despite their smaller magnitude, and that practical longitudinal monitoring will depend on positioning correction, frequency and channel selection, and noise-aware detectability analysis.
\end{abstract}
\section{Introduction}

Neoadjuvant chemotherapy can substantially reduce breast tumors before
surgery. An imaging method that can be repeated during treatment and can detect response early could help identify patients who are not responding as expected. Microwave imaging (MWI) is attractive for this purpose because it is non-ionizing and sensitive to the dielectric properties of breast tissue \cite{lazebnik2007large}, and an initial clinical study has reported treatment-related changes in reconstructed dielectric properties during chemotherapy \cite{meaney2013microwave}. 

A major difficulty, however, is that microwave signals measured at different treatment visits may differ due to factors beyond tumor changes alone. The tumor may shrink or its dielectric properties
may change, but the breast may also be positioned differently, the
amount and distribution of fibroglandular tissue (FGT) may differ in the
modeled volume, and the measurement geometry may change. If these
effects produce larger changes in the measured microwave signals
(scattering parameters, or S-parameters) than the tumor-related changes,
simply comparing measurements between treatment visits may give a
misleading picture of treatment response.

Addressing this problem requires separating four questions that are easily conflated. \textbf{Sensitivity} is the magnitude of the change in the measured S-parameters produced by a change in a model parameter, such as tumor permittivity or breast position. \textbf{Detectability} is whether that change exceeds measurement noise and uncertainty. \textbf{Distinguishability} is whether the multichannel sensitivity pattern of one parameter differs from that of another, quantified here by the cosine similarity between tumor permittivity and other Jacobian columns. \textbf{Identifiability} is whether the full set of parameter effects is linearly independent, quantified here by the singular values of the column-normalized, multi-frequency Jacobian. A tumor-related change may be large yet poorly distinguishable from other effects (e.g., an FGT change), or distinct in pattern yet too small to detect; the two failures call for different remedies.

In this work we build a patient-specific computational framework to quantify sensitivity, distinguishability, and identifiability of tumor dielectric changes relative to FGT permittivity changes and small breast-positioning changes in longitudinal microwave breast imaging.

\section{Materials and Methods}

\subsection{I-SPY 1 imaging data and study time
points}

Dynamic contrast-enhanced magnetic resonance imaging (DCE-MRI) data were obtained from the public I-SPY 1/ACRIN 6657 collection
in The Cancer Imaging Archive (TCIA)\cite{newitt9multi, chitalia2022expert}. The collection
contains longitudinal breast MRI acquired during neoadjuvant
chemotherapy. In the ACRIN 6657 protocol, T1 was acquired within four
weeks before the start of anthracycline-cyclophosphamide chemotherapy;
T2 was acquired at least two weeks after the first cycle and before the
second cycle; T3 was acquired between anthracycline-cyclophosphamide
treatment and taxane therapy when taxane was given; and T4 was acquired
after the final chemotherapy treatment and before surgery. The original
protocol used 1.5-T breast MRI with a dedicated breast coil and a
contrast-enhanced T1-weighted series acquired before and after contrast
administration \cite{hylton2016neoadjuvant}.

Two patients, I-SPY 1-1041 and I-SPY 1-1220, were selected for this computational analysis because both had usable T1-T4 imaging and represented substantially different tumor-volume ranges (Table~\ref{tab:volumes}).

\subsection{Breast, FGT, and tumor
segmentation}

For each patient and treatment time point, the pre-contrast, early post-contrast, and late post-contrast DCE-MRI volumes were assembled from the original Digital Imaging and
Communications in Medicine (DICOM) images and saved in NIfTI format on a common voxel grid.

\begin{figure}[t]
    \centering

    \includegraphics[width=0.93\columnwidth, trim=0 2.6cm 0 1.6cm, clip]{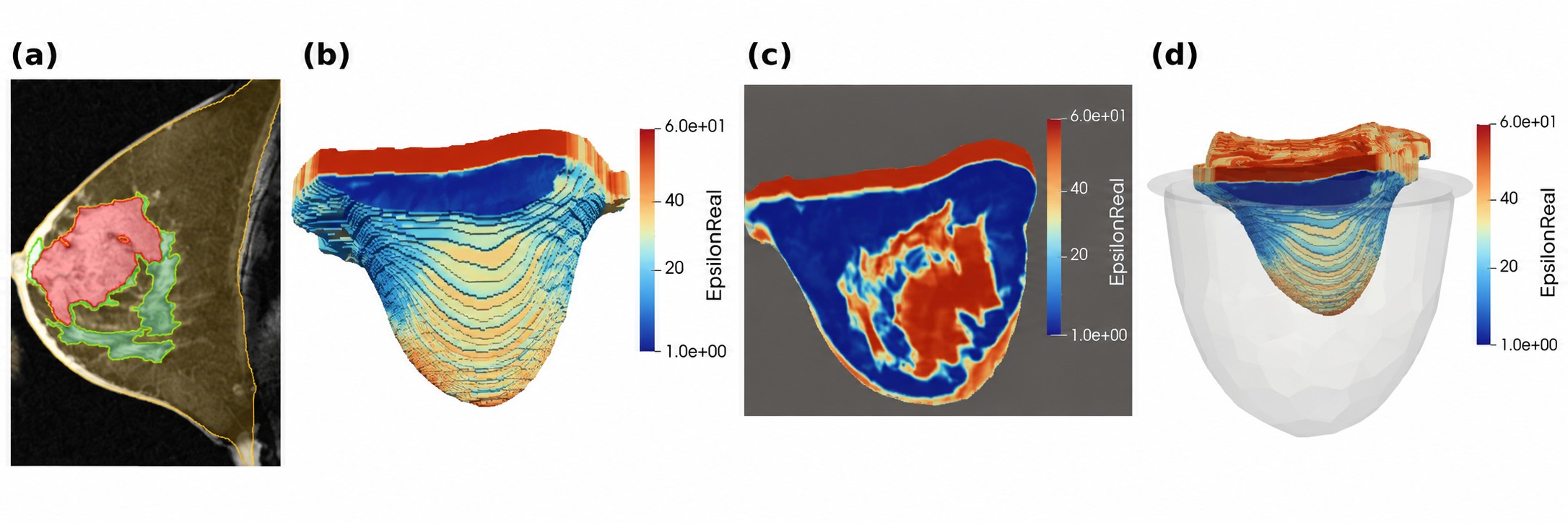}
    \caption{Patient-specific breast model preparation for MWI simulations.
    (a) MRI-based segmentation of breast tissue, tumor, and FGT.
    (b) Three-dimensional dielectric breast model. A 2 mm skin layer and a 10 mm posterior muscle layer were incorporated during model preparation. 
    (c) Cross-sectional distribution of assigned real relative permittivity. 
    (d) Breast model positioned within the actual finite-element imaging chamber, with perfect electric conductor (PEC) surfaces shown in transparent gray.
    }
    \label{fig:preparation}
    \vspace{-10pt}

\end{figure}

Breast and FGT were segmented from the pre-contrast volume with the publicly released BAMF/AIMI Breast-MR nnU-Net~v2 model (Dataset009\_Breast; Zenodo doi:10.5281/zenodo.11998679) \cite{isensee2021nnu, murugesan2025ai}. The model output labels were background, breast/fat, and FGT (Fig.~\ref{fig:preparation}(a)). 

Tumor was segmented from the early post-contrast volume with the nnU-Net baseline released with the MAMA-MIA dataset \cite{garrucho2025large}. The predicted tumor segmentation was subsequently  mapped back to the original pre-contrast MRI grid. All segmentations were visually inspected and used without manual correction.

\subsection{Tissue volumes}

Tumor, FGT, and whole-breast volumes (the union of the fat, FGT, and tumor labels) were computed from the final masks as the labeled voxel count multiplied by the voxel volume (Table~\ref{tab:volumes}). Whole-breast and FGT volumes varied across treatment visits, with particularly large non-monotonic FGT variation in patient 1041 (30.7–62.1 mL). These differences may reflect variations in MRI acquisition, positioning in the breast coil, and automatic segmentation, as well as anatomical change. This between-visit variability is itself a potential confound for longitudinal MWI measurements, as discussed in Section~\ref{sec:limitations}.

\begin{table}[t]
\centering\caption{Tumor, FGT, and breast volumes across four treatment time points}
\label{tab:volumes}\scriptsize
\resizebox{\columnwidth}{!}{\begin{tabular}{llrrrr}
\toprule
Patient & Time & Tumor (mL) & FGT (mL) & Breast (mL) & Tumor/breast \\
\midrule
1041 & T1 & 97.475 & 62.140 & 926.243 & 10.52\% \\
1041 & T2 & 72.284 & 56.012 & 1067.060 & 6.77\% \\
1041 & T3 & 31.653 & 30.729 & 846.022 & 3.74\% \\
1041 & T4 & 20.248 & 47.306 & 830.436 & 2.44\% \\
1220 & T1 & 9.204 & 54.512 & 1618.279 & 0.57\% \\
1220 & T2 & 5.644 & 60.810 & 1578.240 & 0.36\% \\
1220 & T3 & 3.685 & 56.792 & 1518.256 & 0.24\% \\
1220 & T4 & 2.766 & 68.307 & 1686.262 & 0.16\% \\
\bottomrule
\end{tabular}}
    \vspace{-8pt}

\end{table}

\subsection{Patient-specific dielectric
models}

The segmented MRI volumes were converted into spatial maps of complex
permittivity using a segmentation-guided, intensity-based dielectric
mapping approach adapted from Zastrow et al.\cite{zastrow2008development}. Fat and
FGT were assigned frequency-dependent dielectric
properties by interpolating between published Cole--Cole breast-tissue
curves \cite{lazebnik2007large} (Fig.~\ref{fig:preparation}(b-c)). Tumor dielectric properties were assigned using
published malignant-tissue Cole--Cole parameters, with interpolation
between the 25th and 75th percentile curves according to tumor MRI
intensity. A 2-mm skin layer and a
10-mm muscle layer were added to approximate the breast surface and
underlying chest-wall anatomy. Skin and muscle dielectric properties were represented using
the parametric models of Gabriel et al. \cite{gabriel1996dielectric}.

\subsection{Microwave Chamber and Finite-Element Method (FEM) Forward Model}

Finite-element electromagnetic field simulations were performed of the numerical dielectric breast models situated within the quasi-resonant flat-faceted MWI chamber described in \cite{AmerZakaria, Nemez2017, Asefi2019}. Each patient-specific dielectric model was incorporated into a
numerical model of the three-dimensional microwave-imaging chamber schematically depicted in Fig.~\ref{fig:preparation}(d). The 24-port S-parameters at the 24 monopole antenna ports of the system were extracted over a frequency range of 1.0--3.0
GHz in 0.2-GHz increments using the method described in  \cite{Lucas}. 

The patient-specific spatial distributions of
complex relative permittivity were introduced into the
solver as voxel-based constitutive-property maps that were interpolated onto a fixed phantom region. The chamber and antenna arrangement also remained fixed across all simulations.  The FEM mesh at the top of the electrically conducting chamber was terminated using absorbing boundary conditions applied to an outer
hemispherical computational boundary having a 140-mm radius. A mesh having a 3 mm average side-length was used in the breast-phantom region
whereas a  7 mm side-length was used in the surrounding computational domain.

\subsection{Analysis Metrics}
The extracted S-parameters were used to perform sensitivity and identifiability analyses. Six breast-model parameters were studied: tumor relative permittivity
(TumorEps), tumor dielectric loss (TumorLoss), FGT
relative permittivity (FGTEps), breast rotation, lateral translation,
and superior-inferior (SI) translation. Tumor and FGT dielectric parameters were perturbed by
$\pm$10\%, breast rotation by $\pm$2$^\circ$, and lateral and SI position by $\pm$1 mm. For each
perturbed parameter, $\theta$,  central finite-difference approximations were
used  for the derivative with respect to the perturbations:

\begin{equation}\mathbf J_k(f)\approx\frac{\mathbf S(f,\boldsymbol\theta+h_k\mathbf e_k)-\mathbf S(f,\boldsymbol\theta-h_k\mathbf e_k)}{2h_k}\label{eq:jacobian}\end{equation}

Here, S is the vectorized complex S-parameter matrix, $\theta$ is the vector of
six model parameters, \(k\) indexes the parameter being perturbed,
\(h_{k}\) is the perturbation step for parameter \(k\), and \(e_{k}\) is
the corresponding unit vector. 

\subsubsection{Absolute and baseline-normalized
sensitivity}

To quantify and compare the effects of tumor-related and positioning-related perturbations on the simulated microwave measurements, we evaluated baseline-normalized S-parameter changes and Jacobian column norms. Jacobian column norms characterize the sensitivity of the microwave response to each parameter. Because the parameters have different physical units, baseline-normalized S-parameter changes were calculated for the specified positive and negative perturbation steps to compare their resulting measurement responses.

\begin{equation}R_{k,\pm}(f)=\frac{\|\mathbf S_{k,\pm}(f)-\mathbf S_0(f)\|_2}{\|\mathbf S_0(f)\|_2}\label{eq:relative}\end{equation}

Here, \(S_{0}\) is the baseline S-parameter vector, and \(S_{k, \pm}\)
represents the response to the positive or negative perturbation of
parameter \(k\).

\subsubsection{Cosine similarity}

To assess the distinguishability of parameter effects, cosine similarity was calculated between the tumor permittivity Jacobian column and each of the other parameter columns at each frequency. The absolute cosine similarity quantifies the alignment of their multichannel sensitivity patterns, independent of their magnitudes. Absolute cosine similarity ranges from 0 to 1, with
values near 0 indicating approximately orthogonal sensitivity patterns
and values near 1 indicating strongly aligned or anti-aligned patterns. However, distinguishability does not guarantee detectability, as even distinct sensitivity patterns may be too weak to detect in the presence of measurement noise.

Complex Jacobian columns were represented by concatenating their real and imaginary components before calculating cosine similarity. 

\begin{equation}C_{tk}(f)=\frac{|\widetilde{\mathbf J}_t(f)^T\widetilde{\mathbf J}_k(f)|}{\|\widetilde{\mathbf J}_t(f)\|_2\|\widetilde{\mathbf J}_k(f)\|_2}\label{eq:cosine}\end{equation}

where

\[{\widetilde{J}}_{k}(f) = \left\lbrack \begin{array}{r}
Re\left( J_{k}(f) \right) \\
Im\left( J_{k}(f) \right)
\end{array} \right\rbrack\]
and \(t\) denotes tumor permittivity.

\subsubsection{Singular-value analysis}
To assess the identifiability of the six parameters, singular-value decomposition (SVD) was performed. At each of the 11 frequencies (1.0–3.0 GHz), the complex Jacobian was converted to a real-valued matrix by vertically concatenating its real and imaginary components. These real-valued Jacobians were then stacked across frequencies to form the multi-frequency Jacobian, denoted by $J_{\mathrm{all}}$, with six columns corresponding to the six parameters. Each column was normalized to unit Euclidean norm before applying SVD:

\begin{equation}\widehat{\mathbf J}_{\mathrm{all}}=\left[\frac{\mathbf J_{\mathrm{all},1}}{\|\mathbf J_{\mathrm{all},1}\|_2}\ \cdots\ \frac{\mathbf J_{\mathrm{all},6}}{\|\mathbf J_{\mathrm{all},6}\|_2}\right]=\mathbf U\boldsymbol\Sigma\mathbf V^T\label{eq:svd}\end{equation}

Here, $J_{\mathrm{all},k}$ denotes the $k$-th column
of the frequency-stacked, real-augmented Jacobian.
The smallest singular value, $\sigma_{\min}$, was
obtained from the diagonal matrix $\Sigma$, and the
condition number was calculated as
$\kappa=\sigma_{\max}/\sigma_{\min}$.
The corresponding right singular vector,
$v_{\min}$, was extracted from $V$.
A nonzero $\sigma_{\min}$ indicates linear
independence of the six parameter effects, while
a smaller $\kappa$ indicates better conditioning.
The largest absolute coefficients in $v_{\min}$
identify the main contributors to the parameter
combination producing the smallest linearized
response in the normalized sensitivity space.

\section{Results and Discussion}

\subsection{Longitudinal anatomy and
sensitivity}

Both patients showed substantial reductions in segmented tumor volume from T1 to T4 (79\% and 70\%) despite a tenfold difference in baseline tumor volume (Table~\ref{tab:volumes}), allowing the same sensitivity analysis to be examined in a large-tumor and a small-tumor case.

\begin{figure}[t]
\centering
\noindent
\makebox[0.05\columnwidth][l]{(a)}
\includegraphics[width=0.8\columnwidth, trim=0 0.9cm 0 0.3cm, clip]{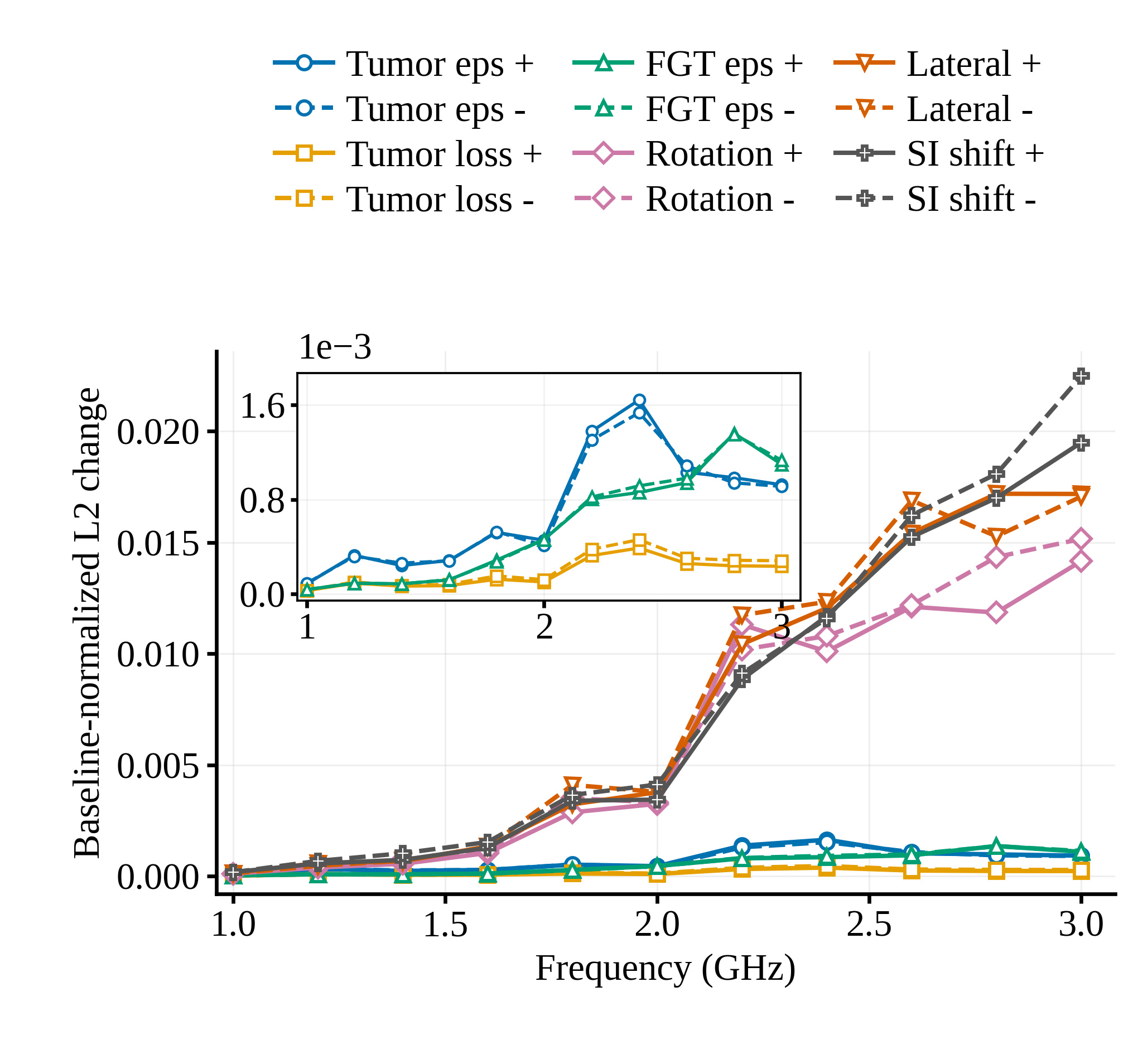}

\noindent
\makebox[0.05\columnwidth][l]{(b)}
\includegraphics[width=0.8\columnwidth, trim=0 0.3cm 0 2.4cm, clip]{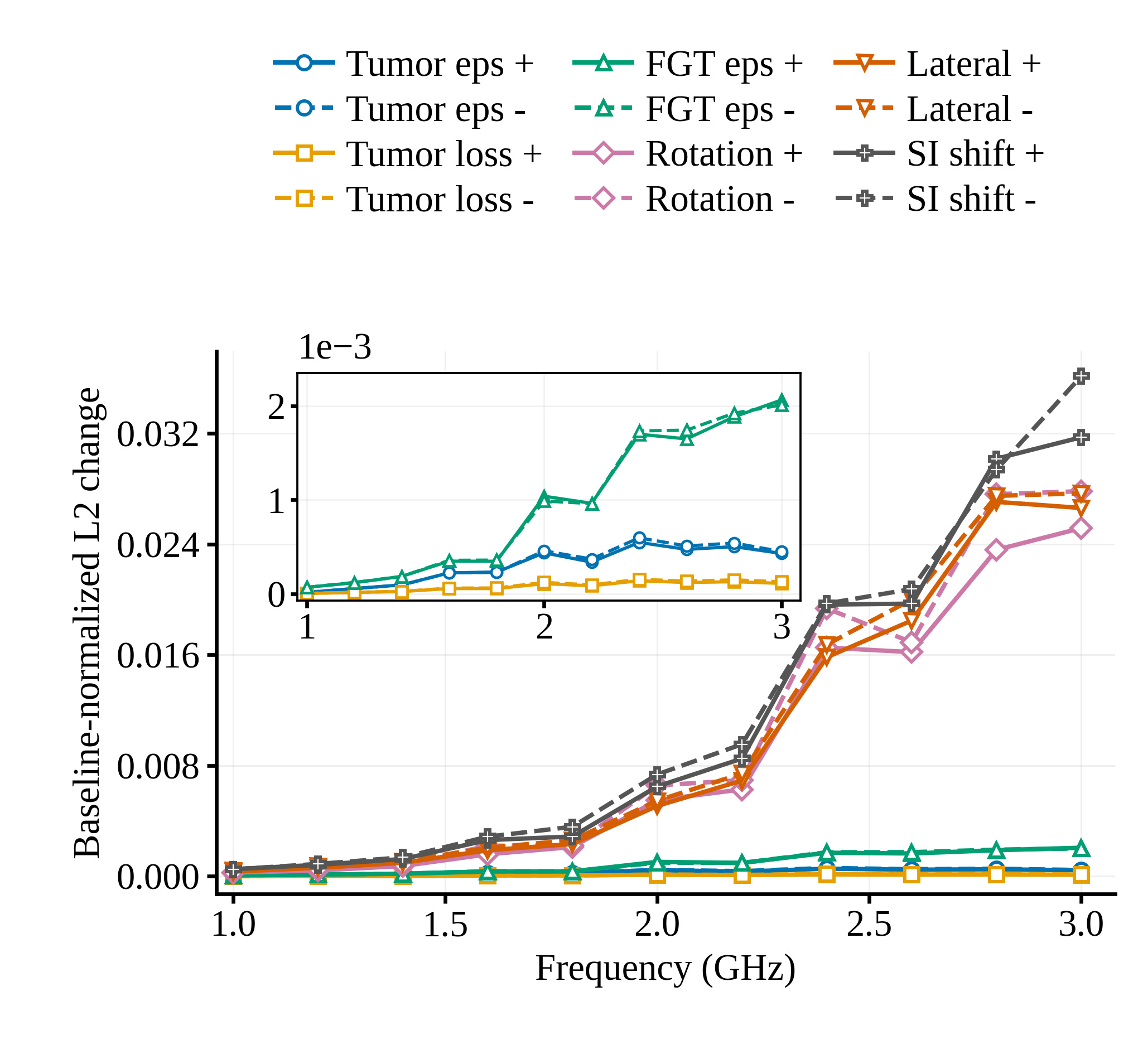}
\caption{Baseline-normalized S-parameter changes versus frequency for (a) I-SPY 1-1041 at T1 and (b) I-SPY 1-1220 at T1, resulting from perturbations in tumor dielectric properties, FGT permittivity, and breast positioning. Solid and dashed curves represent positive and negative perturbations, respectively. Magnified views highlight the tumor- and FGT-related responses.}\label{fig:sensitivity}
\vspace{-8pt}

\end{figure}

At the tested perturbation magnitudes ($\pm$1~mm, $\pm$2$^\circ$, $\pm$10\%), positioning perturbations generally produced larger baseline-normalized S-parameter changes than tumor dielectric perturbations (Fig.~\ref{fig:sensitivity}). For the two T1 cases shown, the relative magnitudes were strongly frequency dependent: below 1.6~GHz, lateral and SI translations generally exceeded the tumor permittivity response by only a factor of about 1--3, whereas above 2~GHz the tumor permittivity response remained on the order of $1\times10^{-3}$ (1041) and $0.5\times10^{-3}$ (1220), while positioning responses continued to rise, becoming approximately one to two orders of magnitude larger at 3~GHz. Tumor loss responses were smaller than tumor permittivity responses. The tumor permittivity response in the small-tumor patient was approximately half that in the large-tumor patient despite a tenfold difference in baseline tumor volume, suggesting that S-parameter sensitivity is not determined by tumor volume alone. The FGT response was also larger relative to the tumor response in 1220 than in 1041, which may partly reflect differences in their FGT-to-tumor volume ratios.

\subsection{Parameter similarity and identifiability}

\begin{figure}[t]
\centering
\noindent
\makebox[0.05\columnwidth][l]{(a)}
\includegraphics[width=0.8\columnwidth, trim=0 0.8cm 0 0.7cm, clip]{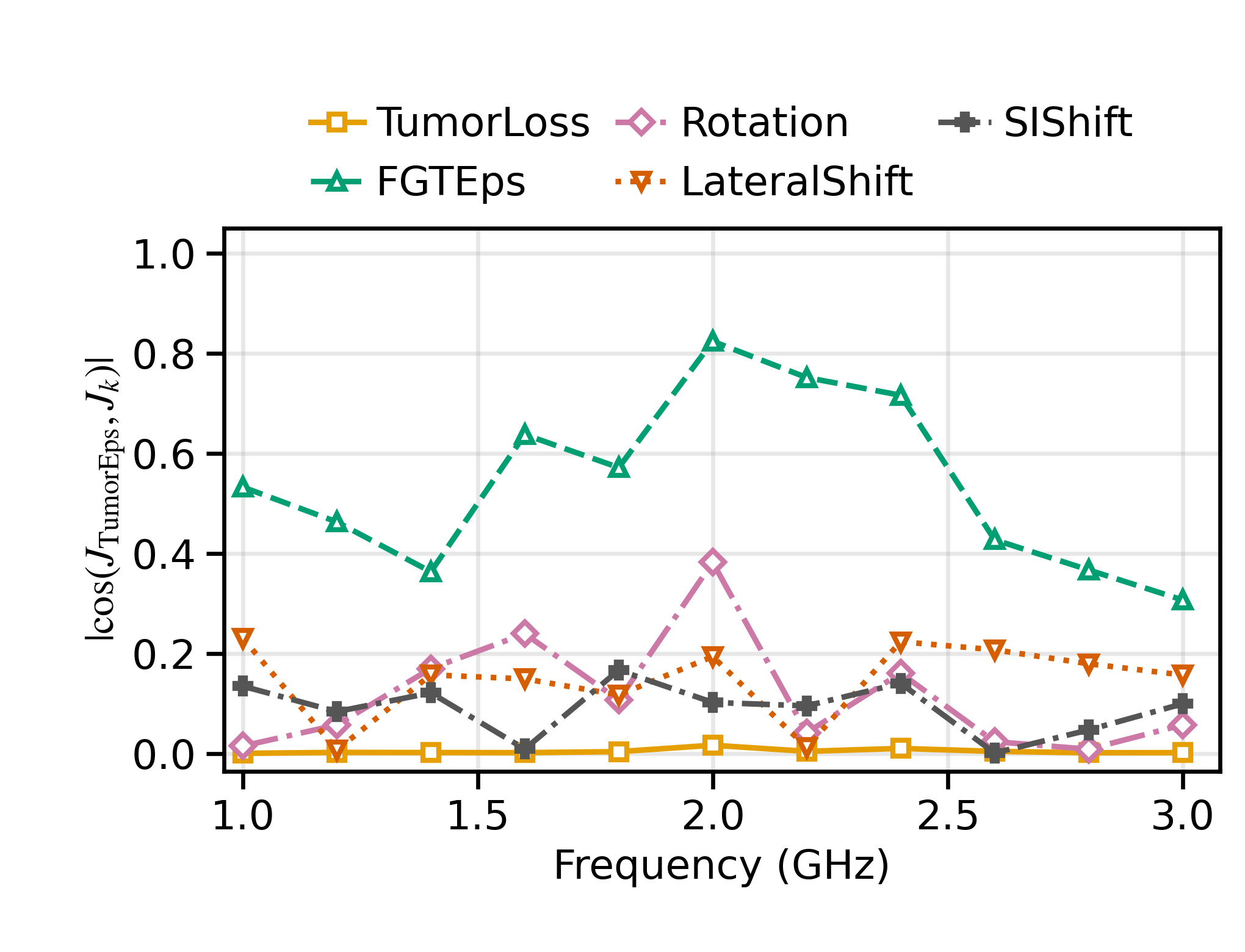}

\noindent
\makebox[0.05\columnwidth][l]{(b)}
\includegraphics[width=0.8\columnwidth, trim=0 0.3cm 0 1.5cm, clip]{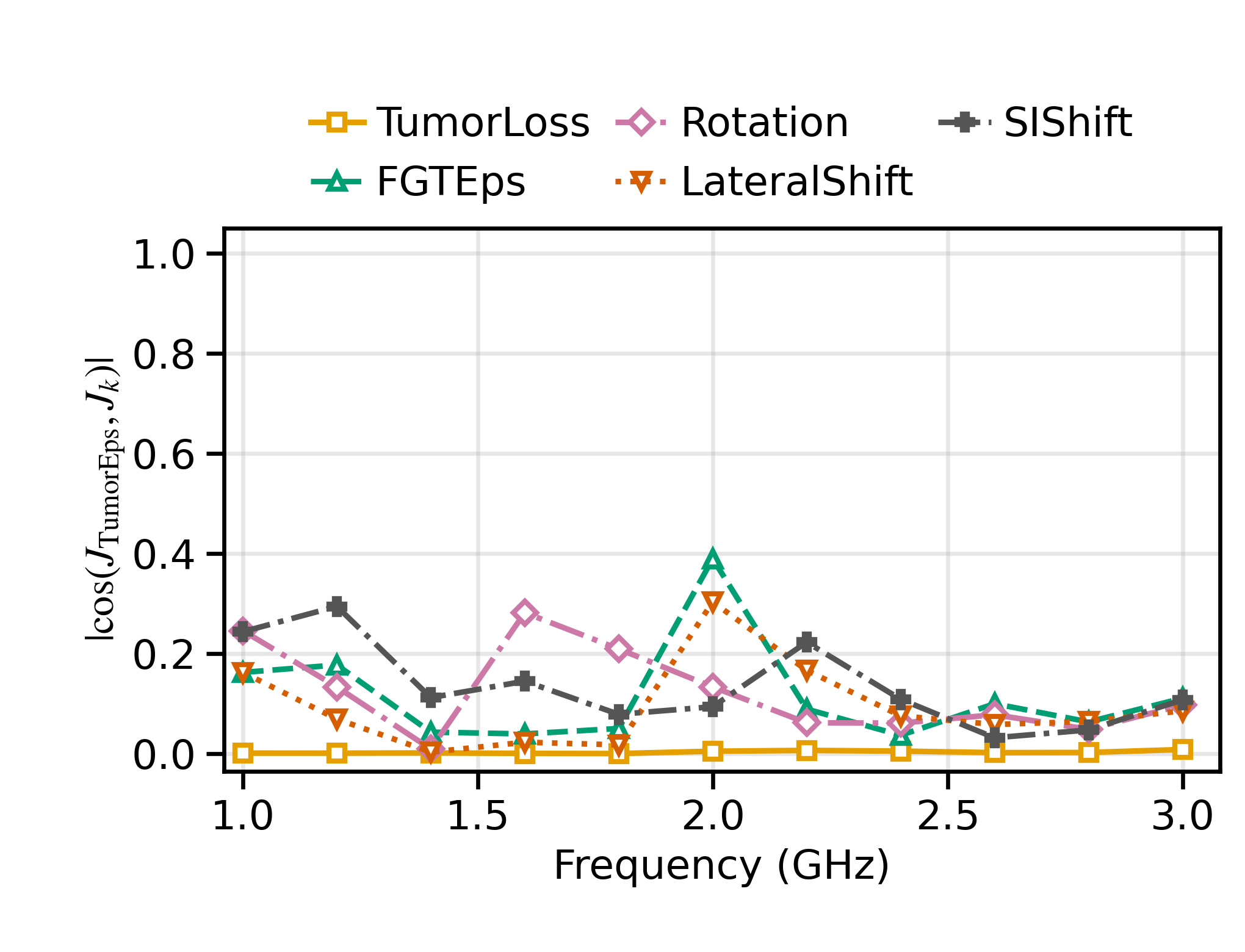}
\caption{Absolute cosine similarity between the tumor permittivity Jacobian column and the other parameter columns as a function of frequency for (a) I-SPY 1-1041 T2 and (b) I-SPY 1-1220 T2. The results illustrate frequency-dependent similarity between tumor and FGT responses, while positioning-related responses are generally less aligned with the tumor permittivity response.}\label{fig:cosine}
\vspace{-8pt}

\end{figure}

Fig.~\ref{fig:cosine} shows that tumor permittivity sensitivity patterns were generally weakly aligned with all three positioning-related sensitivities, suggesting that tumor dielectric changes produce distinguishable multichannel response patterns despite the larger measurement changes caused by positioning perturbations. 
In contrast, tumor--FGT permittivity similarity varied substantially across patients, treatment visits, and frequencies. For I-SPY~1-1041, the maximum absolute cosine similarity was approximately 0.71, 0.82, 0.70, and 0.60 at T1--T4, respectively, with the highest value occurring near 2.0~GHz at T2. For I-SPY~1-1220, the corresponding maxima were approximately 0.23, 0.39, 0.34, and 0.79, with the highest value occurring at 1.0~GHz at T4. At T2, the substantially greater similarity in 1041 (0.82 versus 0.39) may be associated with patient-specific anatomy, including differences in tumor volume (72.3~mL versus 5.6~mL), location, and FGT distribution. However, the elevated similarity in 1220 at T4 suggests that tumor volume alone does not determine tumor--FGT sensitivity-pattern alignment. These findings suggest that the frequencies associated with tumor--FGT ambiguity are patient- and visit-dependent, motivating further investigation of patient-specific frequency selection.

Tumor permittivity and tumor loss sensitivities exhibited near-zero absolute cosine similarity across all eight models (maximum values ranging from 0.0025 to 0.0211). This is consistent with the approximately orthogonal real-augmented sensitivity patterns expected when perturbations of the real and imaginary components of tumor permittivity produce responses differing primarily in phase. Consequently, the low similarity should be interpreted in the context of the complex-response representation rather than as independent evidence of practical parameter recoverability.

\begin{table*}[!t]
\centering
\caption{Singular-value analysis and absolute coefficients
of the smallest right singular vector}
\label{tab:svd}
\setlength{\tabcolsep}{5pt}
\renewcommand{\arraystretch}{1.08}
\begin{tabular}{cc cc cccccc}
\hline
patient & Time & $\sigma_{\min}$ & $\kappa$
& Tumor $\varepsilon$ & Tumor loss & FGT $\varepsilon$
& Rotation & Lateral shift & SI shift \\
\hline
1041 & T1 & 0.72 & 1.65 & 0.14 & 0.03 & 0.14 & \textbf{0.66} & 0.31 & 0.65 \\
1041 & T2 & 0.69 & 1.79 & \textbf{0.69} & 0.07 & 0.68 & 0.04 & 0.18 & 0.12 \\
1041 & T3 & 0.78 & 1.56 & 0.63 & 0.18 & \textbf{0.65} & 0.18 & 0.15 & 0.30 \\
1041 & T4 & 0.51 & 2.50 & 0.11 & 0.05 & 0.08 & 0.58 & 0.50 & \textbf{0.62} \\
1220 & T1 & 0.76 & 1.64 & 0.16 & 0.05 & 0.02 & \textbf{0.72} & 0.16 & 0.65 \\
1220 & T2 & 0.69 & 1.79 & 0.05 & 0.13 & 0.06 & \textbf{0.70} & 0.58 & 0.37 \\
1220 & T3 & 0.78 & 1.47 & 0.13 & 0.10 & 0.05 & \textbf{0.67} & 0.46 & 0.54 \\
1220 & T4 & 0.72 & 1.72 & 0.02 & 0.06 & 0.10 & \textbf{0.72} & 0.47 & 0.49 \\
\hline
\end{tabular}
\parbox{0.95\textwidth}{\footnotesize
$\sigma_{\min}$ is the smallest singular value of the
column-normalized, multi-frequency Jacobian;
$\kappa=\sigma_{\max}/\sigma_{\min}$ is its condition number.
The six rightmost columns report $|v_{\min,k}|$.
Bold values indicate the largest absolute coefficient
for each case.}
\vspace{-10pt}

\end{table*}

The SVD results in Table~\ref{tab:svd} are consistent with the cosine-similarity analysis. The column-normalized multi-frequency Jacobians had smallest singular values of 0.51--0.78 and condition numbers of 1.47--2.50 across the eight models, indicating full column rank and relatively good conditioning within the modeled six-parameter space. Because column normalization removes differences in sensitivity magnitude, these results support linearized identifiability within the modeled six-parameter space but do not establish reliable parameter recovery under measurement noise.

The composition of the weakest right singular vector $\mathbf v_{\min}$ varied across patients and treatment visits. For I-SPY~1-1041 at T2 and T3, tumor and FGT permittivity dominated $\mathbf v_{\min}$ (absolute coefficients 0.69/0.68 and 0.63/0.65, respectively), consistent with their elevated cosine similarity at selected frequencies. This indicates that particular combinations of tumor- and FGT-permittivity changes produce comparatively smaller linearized responses in normalized sensitivity space. 
At T1 and T4 for I-SPY~1-1041, and at all four time points for I-SPY~1-1220, the weakest directions were instead dominated by positioning parameters. These findings suggest that combinations of rotation and translation may partially compensate for one another in the simulated responses. The highest condition number occurred for I-SPY~1-1041 at T4 ($\kappa\approx2.50$), although the Jacobian remained well-conditioned within the modeled parameter space.

\subsection{Implications and limitations}
\label{sec:limitations}
The results distinguish two aspects of longitudinal MWI: the magnitude of tumor-related measurement changes and their distinguishability from competing parameter effects. A 10\% tumor permittivity perturbation produced baseline-normalized S-parameter changes on the order of $10^{-3}$ in the illustrated T1 cases, highlighting the importance of measurement repeatability and calibration accuracy a system must achieve between visits. At frequencies above 2~GHz, the tested positioning perturbations ($\pm$1~mm and $\pm$2$^\circ$) generally produced substantially larger responses, emphasizing the need to investigate repeatable positioning and correction methods. However, the cosine-similarity and SVD analyses indicate that larger positioning responses do not necessarily imply mathematical indistinguishability from tumor-related changes. The observed frequency dependence also motivates investigation of patient-specific frequency selection to balance tumor sensitivity and ambiguity with positioning and FGT effects.

Several limitations qualify these findings. The relative sensitivity magnitudes depend on the chosen perturbation sizes, and the Jacobians characterize responses around each visit-specific baseline rather than directly measuring the effects of longitudinal tumor shrinkage. Although a consistent nominal mesh resolution was used across models, discretization and interpolation may influence the computed positioning sensitivities; mesh-convergence analysis would be needed to assess these effects. MRI-derived whole-breast and FGT volumes also varied across visits, potentially reflecting differences in image coverage, positioning, segmentation, or anatomy, which may introduce additional uncertainty into longitudinal comparisons. Furthermore, favorable conditioning of the column-normalized Jacobians supports linearized identifiability within the modeled six-parameter space but does not establish practical detectability under measurement noise, calibration errors, or model mismatch. The findings are also limited to two patients and one chamber configuration. Future work should evaluate longitudinal S-parameter differences directly, incorporate measurement-noise models, and investigate positioning-correction and frequency-selection strategies.

\section{Conclusion}
We assessed the sensitivity, distinguishability, and linearized identifiability of tumor-related dielectric changes using eight MRI-derived breast models from two I-SPY~1 patients. Tumor-permittivity perturbations produced relatively small S-parameter changes compared with positioning perturbations, particularly above 2~GHz. Nevertheless, tumor permittivity sensitivity patterns were generally distinguishable from positioning-related patterns, and the column-normalized multi-frequency Jacobians remained well-conditioned ($\kappa\leq2.5$). Tumor--FGT similarity varied across patients, visits, and frequencies. These findings demonstrate that response magnitude and parameter distinguishability are distinct considerations: larger positioning responses do not necessarily make tumor-related effects mathematically indistinguishable. Practical longitudinal monitoring will require further evaluation of measurement noise, positioning repeatability and correction, and patient-specific frequency selection to establish whether chemotherapy-related changes can be reliably detected.

\bibliographystyle{IEEEtran}
\bibliography{IEEEabrv,bibliography}

\end{document}